\documentclass[twocolumn,showpacs,preprintnumbers,amsmath,amssymb]{revtex4-1}

\usepackage{graphicx}% Include figure files
\usepackage{dcolumn}% Align table columns on decimal point
\usepackage{hyperref}
\usepackage{floatrow}
\usepackage{amsmath}
\usepackage{epsfig}
\usepackage{subfig}
\usepackage{array}
\usepackage{xcolor}
\usepackage{bm}% bold math

\def\be{\begin{equation}}
\def\ee{\end{equation}}
\def\bea{\begin{eqnarray}}
\def\eea{\end{eqnarray}}
\def\NO{\nonumber}
\def\gev{\mathrm{~GeV}}

\def\md{\mathrm{d}}

\begin{document}

\normalsize

\title{Kinematic distributions of the $\eta_c$ photoproduction in $ep$ collisions within the nonrelativistic QCD framework}

\author{Hong-Fei Zhang}
\email{hfzhang@ihep.ac.cn}
\affiliation{College of Big Data Statistics, Guizhou University of Finance and Economics, Guiyang, 550025, China}
\author{Yu Feng}
\email{yfeng@ihep.ac.cn}
\affiliation{Department of Physics, College of Basic Medical Sciences, Army Medical University, Chongqing, 400038, China}
\author{Wen-Long Sang}
\email{wlsang@ihep.ac.cn}
%\email{Corresponding author: wlsang@ihep.ac.cn}
\affiliation{School of Physical Science and Technology, Southwest University, Chongqing, 400700, China}
\author{Yu-Peng Yan}
\email{yupeng@g.sut.ac.th}
\affiliation{School of Physics and Center of Excellence in High Energy Physics $\mathrm{\&}$ Astrophysics,
Suranaree University of Technology, Nakhon Ratchasima 30000, Thailand}

\begin{abstract}
We study the $\eta_c$ photoproduction in $ep$ collisions in this paper.
The short-distance coefficients for $c\bar{c}(^1S_0^{[1]})$, $c\bar{c}(^1S_0^{[8]})$, $c\bar{c}(^3S_1^{[8]})$,
and $c\bar{c}(^1P_1^{[8]})$ photoproductions are evaluated at leading order in $\alpha_s$ expansion,
where the color-singlet contribution is achieved for the first time.
We have carefully analyzed different kinematic distributions of the cross sections and
found that the color-singlet contribution is considerably suppressed comparing with the color-octet parts.
This feature renders the $\eta_c$ photoproduction process an ideal laboratory to test the color-octet mechanism in nonrelativistic QCD.
By taking different sets of long-distance matrix elements,
we have observed some apparently distinguishable predictions,
which can be utilized to scrutinize the validity of these matrix elements.
\end{abstract}

\maketitle

\section{Introduction\label{sec:introduction}}

In 2014, the $\eta_c$ hadroproduction cross section was measured for the first time by the LHCb Collaboration~\cite{Barsuk:2012ic, Aaij:2014bga},
which opened a window for the study of the pseudoscalar quarkonia production.
Under the assumptions of the heavy quark spin asymmetry,
the $J/\psi$ and $\eta_c$ hadroproduction was studied comprehensively in Refs.~\cite{Butenschoen:2014dra, Han:2014jya, Zhang:2014ybe},
the first one of which considered the $\eta_c$ measurement as a big challenge of nonrelativistic QCD (NRQCD)~\cite{Bodwin:1994jh},
while the latter two reconciled the $J/\psi$ and $\eta_c$ hadroproduction data within the NRQCD framework.
Unfortunately, when making use of the long-distance matrix elements (LDMEs) obtained in Refs.~\cite{Han:2014jya, Zhang:2014ybe} in the $J/\psi$ photoproduction,
the theoretical results significantly overshoot the data~\cite{Butenschoen:2017iks}.
Still, there does not exist any set of LDMEs
which can describe all the data on charmonium production and polarization up to QCD next-to-leading-order (NLO).
For some quarkonium production processes, the higher-order corrections in both $\alpha_s$ and $v^2$ expansion proved to be important,
which might radically change the phenomenological results.
Accordingly, we would expect the universality of the LDMEs be observed when higher precision calculations are completed.
Recently, two-loop calculations for some quarkonium production and decay processes
have been achieved by several theory groups~\cite{Feng:2015uha, Sang:2015uxg, Chen:2017xqd, Feng:2017hlu, Chen:2017pyi, Chen:2017soz, Feng:2019zmt}.
There are also efforts made to find new approaches to improving the precision of the theoretical predictions,
such as improving the convergence quality of the NRQCD expansions~\cite{Ma:2017xno},
and using fragmentation functions to achieve the dominant contributions
at higher orders in high $p_t$ regions~\cite{Bodwin:2014gia, Bodwin:2015yma, Bodwin:2015iua}.

In order to test NRQCD, it is also helpful to find such processes in which the color-octet (CO) mechanism dominates the quarkonium production.
As a perfect example, inclusive $J/\psi$ hadroproduction has been thoroughly investigated at QCD NLO in a series of theoretical papers
\cite{Campbell:2007ws, Ma:2010yw, Butenschoen:2010rq, Ma:2010jj, Butenschoen:2011yh,
Butenschoen:2012px, Chao:2012iv, Gong:2012ug, Shao:2014yta, Sun:2015pia, Feng:2018ukp}.
As another example, the color-singlet (CS) $J/\psi$ photoproduction undershoot the data~\cite{Artoisenet:2009xh},
thus the CO mechanism is needed~\cite{Butenschoen:2009zy, Butenschoen:2011ks}.
The $J/\psi$ leptoproduction is also a good laboratory to test NRQCD.
The QCD corrections to the CS $J/\psi$ leptoproduction has been given in Ref.~\cite{Sun:2017wxk}.
However, QCD NLO results for the CO processes are still lacking.
Interestingly, even the QCD leading-order calculation of this process is not trivial.
The first correct results were obtained as recently as in 2017~\cite{Zhang:2017dia, Sun:2017nly}.

Thanks to its specific quantum numbers,
the CS $\eta_c$ is excluded from many processes at QCD leading order,
and therefore considered to be suppressed.
An interesting example is the $\eta_c$ production in association with light hadrons in $e^+e^-$ annihilation~\cite{Gong:2016jiq},
where the CS contributions are negligible comparing with the CO ones.
The $\eta_c$ and $\eta_c'$ hadroproduction are also studied in greater detail in Refs.~\cite{Lansberg:2017ozx, Feng:2019zmn}.
As early as in 1999, the $\eta_c$ photoproduction~\cite{Hao:1999kq} and leptoproduction~\cite{Hao:2000ci}
in electron-proton ($ep$) collisions has been proposed to test the CO mechanism,
because the CS $\eta_c$ can be produced with at least two gluons emitted,
which is suppressed by a factor of $\alpha_s^2$ comparing to $c\bar{c}(^1S_0^{[8]})$ production.
Therefore the CO contribution is thought to dominate this process.
However the CO LDMEs are suppressed relative to their CS counterpart.
One may wonder whether the CS contribution is dispensable.
Limited by the calculation capability at that time,
the CS photoproduction and leptoproduction were not evaluated in Refs.~\cite{Hao:1999kq, Hao:2000ci}.
Therefore, to make a solid conclusion, we need to complement this part and compare it with the CO contributions.

In this paper, we restudy the $\eta_c$ photoproduction by including both the CS and CO processes.
Section~\ref{sec:framework} gives a brief introduction to the NRQCD framework for the calculation of the $\eta_c$ photoproduction.
In this framework, the numerical results are evaluated and the phenomenological analyses are given in Section~\ref{sec:numerical}.
In Section~\ref{sec:summary}, a brief summary and conclusion is presented.

\section{$\eta_c$ Photoproduction in NRQCD Framework\label{sec:framework}}

When the scattering angle of the electron is very small in $ep$ collisions,
the cross sections can be evaluated approximately by calculating the process of a proton
interacting with a real photon~\cite{vonWeizsacker:1934nji, Williams:1934ad, Frixione:1993yw},
which carries a fraction ($y$) of the momentum of the incident electron.
This picture is known as the Weizs$\mathrm{\ddot{a}}$cker-Williams Approximation (WWA).
Under this approximation, the $\eta_c$ production cross section can be written as
\bea
&&\md\sigma(e(k)+p(P)\rightarrow\eta_c(p)+X(p_X)+e(k')) \NO \\
&&~~~~=\int\md yf_{\gamma/e}(y)\md\sigma_{\gamma p\rightarrow\eta_c+X}(q=yk), \label{eqn:cs}
\eea
where $q$ is the momentum of the incident photon,
and $f_{\gamma/e}$ is the WWA photon distribution function,
the form of which can be found in Ref.~\cite{Frixione:1993yw} and is presented explicitly as follows,
\bea
&&f_{\gamma/e}(y)=\frac{\alpha}{2\pi}\{\frac{2m_e^2y}{Q_{max}^2}-\frac{2(1-y)}{y} \NO \\
&&~~~~~~+\frac{2(1-y)+y^2}{y}\mathrm{ln}\left[\frac{Q_{max}^2}{m_e^2}\frac{(1-y)}{y^2}\right]\}.
\eea
Here $m_e$ is the electron mass, $\alpha$ is the fine structure constant,
and $Q_{max}$ is the maximum value of $Q$, which is determined by the experiment.
$Q$ is defined by
\bea
Q^2=-(k-k')^2. \label{eqn:q2}
\eea
Note that $X$ in Eq. (\ref{eqn:cs}) represents one or several particles.

Within the NRQCD framework, $\eta_c$ can be produced via four intermediate $c\bar{c}$ states,
namely $^1S_0^{[1]}$, $^1S_0^{[8]}$, $^3S_1^{[8]}$, and $^1P_1^{[8]}$, up to the relative order of $v^4$,
where $v$ is the typical relative velocity of the $c$-quark in $\eta_c$.
The cross section for the $\eta_c$ photoproduction can thus be expressed as
\bea
\md\sigma_{\gamma p\rightarrow\eta_c+X}=\sum_n\md\hat{\sigma}_{\gamma p\rightarrow c\bar{c}(n)+X}\langle\mathcal{O}^{\eta_c}(n)\rangle, \label{eqn:csg}
\eea
where $n$ runs over the four intermediate $c\bar{c}$ states,
and $\md\hat{\sigma}_{\gamma p\rightarrow c\bar{c}(n)+X}$ and $\langle\mathcal{O}^{\eta_c}(n)\rangle$
are the corresponding short-distance coefficient (SDC) and LDME, respectively.
The SDCs can be evaluated perturbatively as expansions in the strong coupling, $\alpha_s$,
while the CO LDMEs in Eq. (\ref{eqn:csg}) are obtained through their correspondence with the LDMEs for the $J/\psi$ production.
The relations are given as
\bea
&&\langle\mathcal{O}^{\eta_c}(^1S_0^{[8]})\rangle=\frac{1}{3}\langle\mathcal{O}^{J/\psi}(^3S_1^{[8]})\rangle, \NO \\
&&\langle\mathcal{O}^{\eta_c}(^3S_1^{[8]})\rangle=\langle\mathcal{O}^{J/\psi}(^1S_0^{[8]})\rangle, \NO \\
&&\langle\mathcal{O}^{\eta_c}(^1P_1^{[8]})\rangle=3\langle\mathcal{O}^{J/\psi}(^3P_0^{[8]})\rangle. \label{eqn:LDME}
\eea
The CS LDME can be estimated by the approximate relation between it and the quarkonium wave function at the origin, $R(0)$,
which is written as
\bea
\langle\mathcal{O}^{\eta_c}(^1S_0^{[1]})\rangle\approx\frac{3}{2\pi}\Big|R(0)\Big|^2.
\eea

According to the parton model, the SDCs can be further factorized as
SDCs for the $c\bar{c}$ production via the photon-parton fusion convoluted with the parton distribution functions (PDFs).
Explicitly, we have
\bea
&&\md\hat{\sigma}_{\gamma p\rightarrow\eta_c+X}=\sum_{i, n}\int\md x f_{i/p}(x,\mu_f) \NO \\
&&~~~~\times\md\hat{\sigma}_{\gamma i\rightarrow c\bar{c}(n)+X}(p_i=xP)\langle\mathcal{O}^{\eta_c}(n)\rangle, \label{eqn:csp}
\eea
where $i$ runs over all possible species of partons whose momenta are denoted by $p_i$,
and $\mu_f$ is the factorization scale.
Note that the partonic SDC, $\md\hat{\sigma}_{\gamma i\rightarrow c\bar{c}(n)+X}$,
also depends on $\mu_f$, which cancels the $\mu_f$ dependence of the PDFs.
Then, the SDC, $\md\hat{\sigma}_{\gamma p\rightarrow\eta_c+X}$, is independent of $\mu_f$.
In collinear factorization, to generate an $\eta_c$ with nonzero transverse momentum ($p_t$),
there are only five partonic processes at QCD LO. They are
\bea
&&\gamma+g\rightarrow c\bar{c}(^1S_0^{[1]})+g+g, \NO \\
&&\gamma+g\rightarrow c\bar{c}(^1S_0^{[8]})+g, \NO \\
&&\gamma+q (\bar{q})\rightarrow c\bar{c}(^1S_0^{[8]})+q (\bar{q}), \NO \\
&&\gamma+g\rightarrow c\bar{c}(^3S_1^{[8]})+g, \NO \\
&&\gamma+g\rightarrow c\bar{c}(^1P_1^{[8]})+g, \label{eqn:process}
\eea
where $q$ represents a light quark ($u$, $d$, or $s$).

The SDCs on the right-hand side of Eq. (\ref{eqn:csp}) can be expressed as
\bea
\md\sigma_{\gamma i\rightarrow c\bar{c}(n)+X}=\frac{1}{2sN}\big|\mathcal{M}_{\gamma i\rightarrow c\bar{c}(n)+X}\big|^2\md\Phi, \label{eqn:sdc}
\eea
where $s=(q+p_i)^2$, $N$ synthesizes the spin, color, and symmetry average factors,
$\mathcal{M}_{\gamma i\rightarrow c\bar{c}(n)+X}$ denotes the amplitude for the $c\bar{c}(n)$ production in the partonic process,
and $\md\Phi$ is the phase space which can be written as
\bea
\md\Phi=(2\pi)^4\delta^4(q+p_i-p-p_X)\frac{\md^3p}{(2\pi)^32p_0}\md\Phi_X, \label{eqn:ps}
\eea
where $\md\Phi_X$ encapsulate the phase space for all the hadronic final-state particles other than the $c\bar{c}$.
To accord with the experiment, we express the cross sections in terms of the following two variables,
\bea
&&W^2=(q+P)^2=2q\cdot P=yS, \NO \\
&&z=\frac{P\cdot p}{P\cdot q}=\frac{p_i\cdot p}{p_i\cdot q}, \label{eqn:wz}
\eea
where $S$ is the squared invariant colliding energy of the $ep$ system,
and $z$ is called the elasticity coefficient.

Substituting Eq. (\ref{eqn:csg}), Eq. (\ref{eqn:csp}), and Eq. (\ref{eqn:sdc}) into Eq. (\ref{eqn:cs}),
we obtain the form of the cross section as
\bea
\md\sigma=\sum_{i,n}\int\md x\md yf_{i/g}(x,\mu_f)f_{\gamma/e}(y) \NO \\
\times\frac{1}{2sN}\Big|\mathcal{M}_{\gamma i\rightarrow c\bar{c}(n)+X}\Big|^2\md\Phi. \label{eqn:csm}
\eea
Integrating over $\md x$, replacing $y$ by $W$,
one can further reduce Eq. (\ref{eqn:csm}) into
\bea
&&\md\sigma=\frac{1}{8\pi sSN}\sum_{i,n}\langle\mathcal{O}^{\eta_c}(n)\rangle f_{i/p}(\frac{s}{W^2},\mu_f)f_{\gamma/e}(\frac{W^2}{S}) \NO \\
&&~~~~~~\times\big|\mathcal{M}_{\gamma i\rightarrow c\bar{c}(n)+X}\big|^2\frac{\md W}{W}\frac{\md z}{z(1-z)}\md p_t^2\md\phi_x, \label{eqn:csf}
\eea
where $\md\phi_x$ is defined as follows:
$\md\phi_x=1$ for all the three CO processes, while for the CS process,
\bea
\md\phi_x=\frac{\md s_{ab}}{2\pi}(2\pi)^4\delta^4(p_X-p_a-p_b) \NO \\
\times\frac{\md^3p_a}{(2\pi)^32E_a}\frac{\md^3p_b}{(2\pi)^32E_b}.
\eea
Here, $p_a$ and $p_b$ denote the momenta of the two final-state gluons,
and $s_{ab}\equiv(p_a+p_b)^2$.

Summing over all the processes in Eq. (\ref{eqn:process}),
we obtain the $\eta_c$ photoproduction cross section.

\section{Numerical Results and Phenomenological Analysis\label{sec:numerical}}

The squared amplitudes for the processes listed in Eq. (\ref{eqn:process}) are calculated automatically by the FDC package~\cite{Wang:2004du}.
In our numerical calculations, we adopt the following parameter choices.
$m_e=0.5\times10^{-3}\gev$, $m_c=1.5\gev$, and $\alpha=1/137$.
The renormalization scale ($\mu_r$) and factorization scale are set to be $\mu_r=\mu_f=\sqrt{4m_c^2+p_t^2}$.
Since HERA has stopped its running, we aim at new $ep$ colliders such as the EIC.
While the running energies of such colliders are not known yet,
we set the energies of the incident beams according to the HERA experiment,
namely the energy of the electron beams is $27.5\gev$ and that of the proton beams is $920\gev$.
The photon distribution function is taken from Ref.~\cite{Frixione:1993yw},
where $Q_{max}^2=0.5\gev^2$ is chosen in our numerical calculation.
We employ CTEQ6L1~\cite{Pumplin:2002vw} as the PDF for the protons.
The CS LDME is computed according to
\bea
\langle\mathcal{O}^{\eta_c}(^1S_0^{[1]})\rangle=\frac{3}{2\pi}\big|R(0)\big|^2\approx0.387\gev^3,
\eea
where the value of the wave function at the origin is taken from Ref.~\cite{Eichten:1995ch} as $|R(0)|^2=0.81\gev^3$.
To obtain the value of $\alpha_s$, the one-loop running equation is employed and its value at the $Z_0$-boson mass is set to be $\alpha_s(M_Z)=0.13$.
In this paper, we calculate the differential cross section with respect to $p_t^2$, $W$, and $z$.
For the $p_t^2$ distribution, the ranges of $W$ and $z$ are constrained by $60\gev<W<240\gev$ and $0.3<z<0.6$, respectively.
In the calculation of the $W$ and $z$ distributions, no constraint on the value of $p_t$ is applied.
The $W$ distribution is calculated in the kinematic region, $0.3<z<0.6$,
while the $z$ distribution is calculated in the kinematic region, $60\gev<W<240\gev$.

\begin{figure}
\includegraphics[scale=0.5]{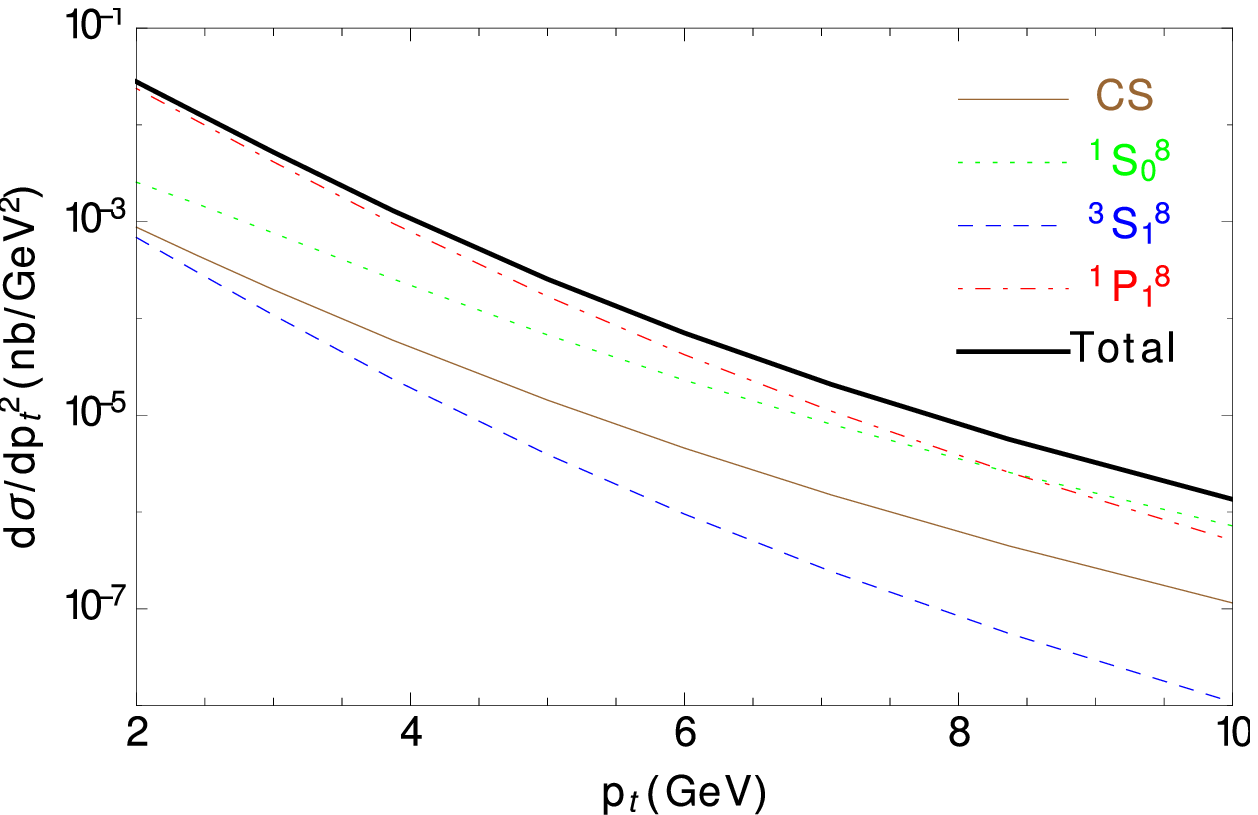}\\
\includegraphics[scale=0.5]{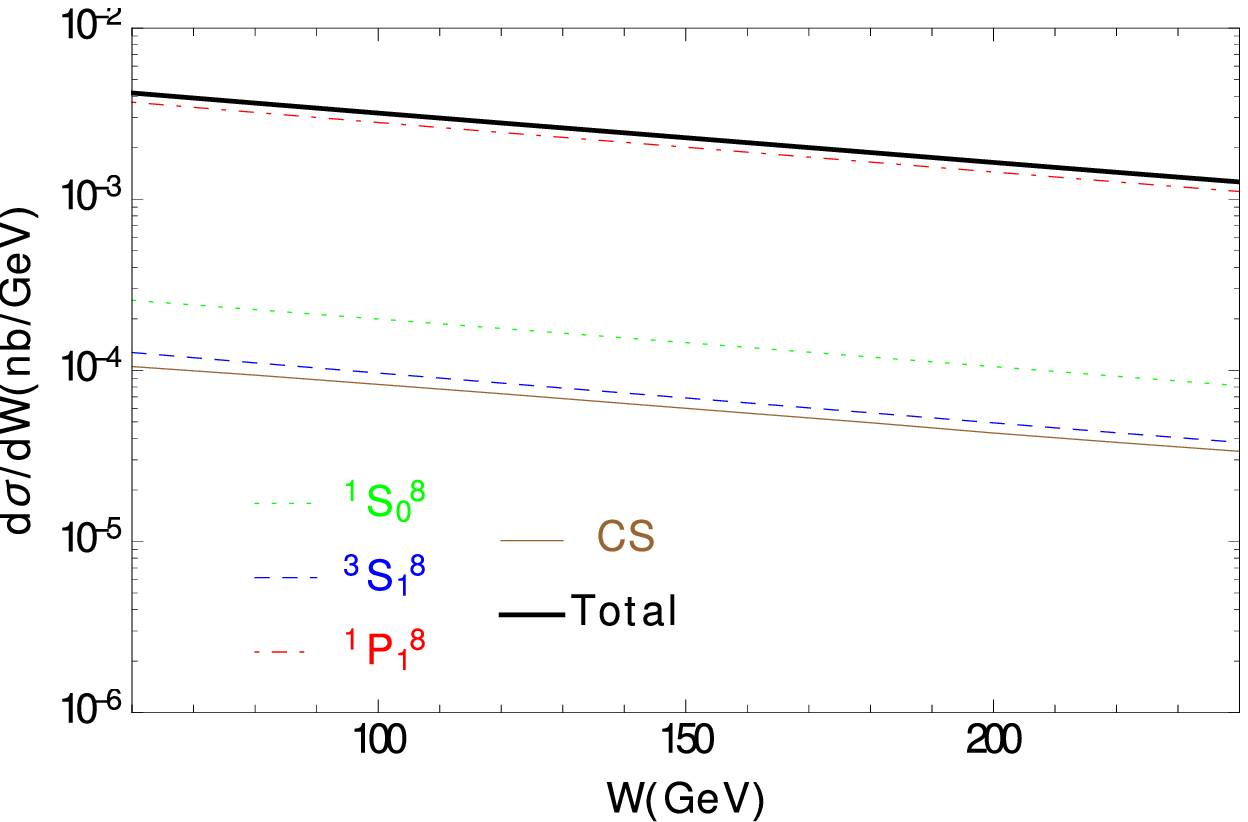}\\
\includegraphics[scale=0.5]{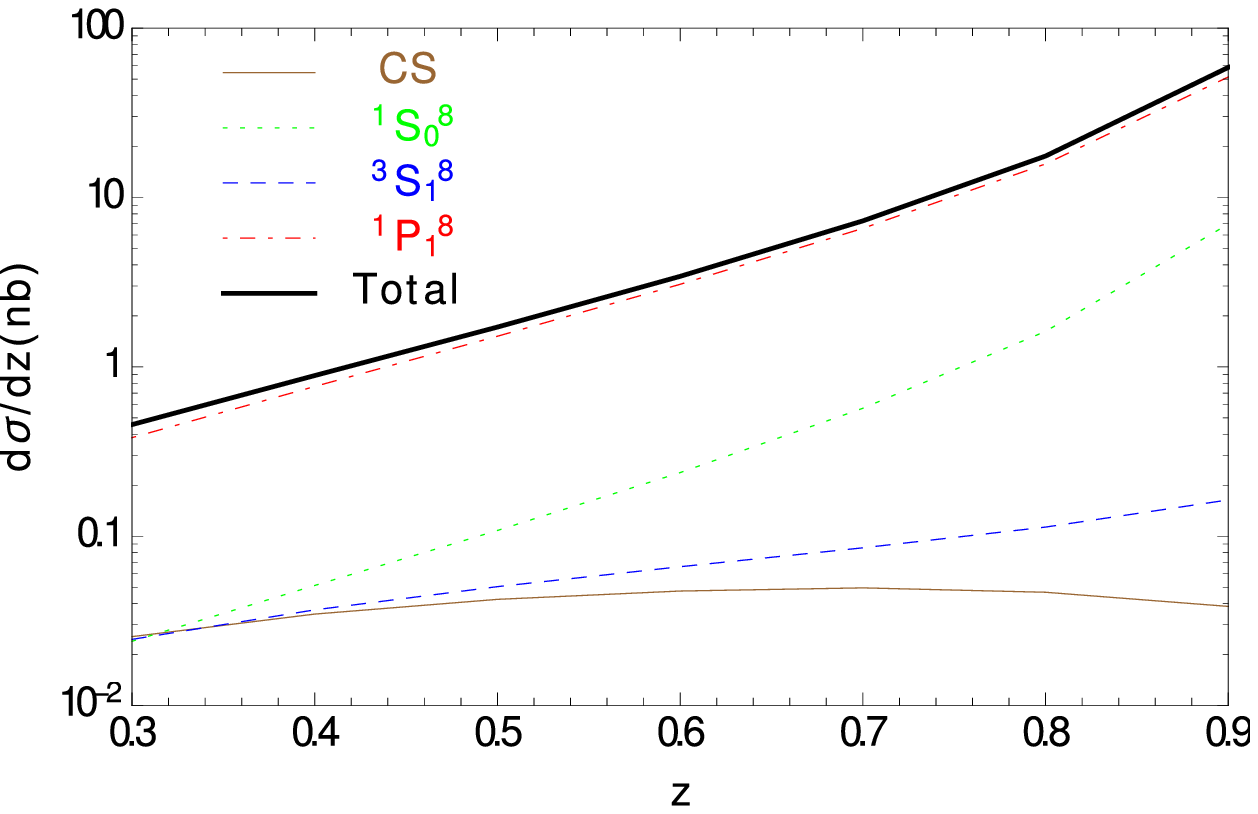}
\caption{\label{fig:co}
The $p_t$ (upper), $W$ (mid), and $z$ (lower) distribution of $\eta_c$ photoproduction.
}
\end{figure}

Fixing the input parameters, we separately calculate the differential cross sections with respect to $p_t^2$, $W$, and $z$.
The numerical results are presented in Fig.~\ref{fig:co},
where the CO LDMEs are taken from Ref.~\cite{Zhang:2014ybe, Sun:2015pia}.
For the sake of convenience, we list these LDMEs below,
\bea
&&\langle\mathcal{O}^{\eta_c}(^1S_0^{[8]})\rangle=0.36\times10^{-2}\gev^3, \NO \\
&&\langle\mathcal{O}^{\eta_c}(^3S_1^{[8]})\rangle=0.74\times10^{-2}\gev^3, \NO \\
&&\langle\mathcal{O}^{\eta_c}(^1P_1^{[8]})\rangle/m_c^2=6.0\times10^{-2}\gev^3, \label{eqn:LDMEz}
\eea
where the relations in Eq. (\ref{eqn:LDME}) have been used.

Although the SDC for CS process is suppressed by a factor of $\alpha_s$ comparing with the CO ones,
the CS LDME is enhanced by some powers of $1/v$.
Since for the charmonia, $v^4$ is approximated to be 0.1,
one cannot naively conclude a CO dominance picture in the $\eta_c$ photoproduction just from the analysis of the scaling.
From Fig.~\ref{fig:co}, we immediately realize that the CS contribution is quite small compared with the CO parts.
This indicates that even the LDME is enhanced a lot,
the SDC of the CS is highly suppressed,
and therefore the CS contribution is negligible.
It is clearly shown that the $^3S_1^{[8]}$ channel is comparable with the CS one,
which is almost negligible.
This is completely different with the $\eta_c$ hadroproduction case,
in which the $^3S_1^{[8]}$ channel dominates the CO $\eta_c$ production.
This feature can be exploited to impose new constraints on the LDMEs.
Our calculation also shows that the cross sections are nearly saturated by the $^1P_1^{[8]}$ channel.
This is partly due to the fact that $^1P_1^{[8]}$ bears a large LDME.
In Fig.~\ref{fig:co}, we also observe a sharp increase as $z$ approaches the endpoint, $z=1$.
We should remember the fact that the perturbative expansion is invalid in the region $z\rightarrow1$,
where the cross sections suffer a divergence, $1/(1-z)$, and a more careful treatment should be implemented,
which however is beyond the scope of this paper.
To efficiently eliminate the effect of this unphysical divergence,
we apply a kinematic cut, $0.3<z<0.6$, in presenting the $p_t^2$ and $W$ distribution of the differential cross sections.
From the lower plot of Fig.~\ref{fig:co}, we can see that in the region $0.3<z<0.6$,
the $z$ distribution of $c\bar{c}(^1S_0^{[8]})$ and $c\bar{c}(^1P_1^{[8]})$ is not so sharp as in the region $z>0.6$,
to this end, we believe that the divergence problem is reasonably mitigated by this cut.
Accordingly, the relative importance of the different channels in the upper and mid plots of Fig.~\ref{fig:co} is reasonable,
which shows remarkable suppression of the CS channel.

Since there are several sets of LDMEs available on the market,
it is necessary and instructive to compare their corresponding predictions for the $\eta_c$ photoproduction and see whether,
just like in the $J/\psi$ hadroproduction case, they work equally well also in the $\eta_c$ photoproduction process.
In the following discussions, we employ five different sets of LDMEs,
taken from Refs.~\cite{Butenschoen:2011yh, Chao:2012iv, Sun:2015pia, Bodwin:2015iua, Feng:2018ukp}, respectively.
They are listed in Table~\ref{tab:ldme},
where the relations in Eq. (\ref{eqn:LDME}) have been implemented.
Each of them is independently obtained, and can describe the $J/\psi$ yield data at the Tevatron and LHC.
Note that the LDMEs for the $J/\psi$ production are extracted also in Refs.~\cite{Gong:2012ug, Bodwin:2014gia},
since the authors of the two references have updated their results in Refs.~\cite{Feng:2018ukp, Bodwin:2015iua}, respectively,
we do not present results for the old version of the LDMEs in this paper.

\begin{table*}
\begin{center}
\caption{
\label{tab:ldme}
The CO LDMEs for the $\eta_c$ production obtained by different theory groups,
where the relations between the LDMEs for the $J/\psi$ and $\eta_c$ production are employed.
}
\begin{tabular}{ccccccc}
\hline
\hline
References&~Butenschon~&~Chao and~&Sun and~&Bodwin and~&Feng and \\
~&and \textit{et al.}~\cite{Butenschoen:2011yh}&~\textit{et al.}~\cite{Chao:2012iv}~&~\textit{et al.}~\cite{Sun:2015pia}~
&~\textit{et al.}~\cite{Bodwin:2015iua}~&~\textit{et al.}~\cite{Feng:2018ukp} \\
\hline
$\langle{\cal O}^{\eta_c}(^1S_0^{[8]})\rangle(10^{-2}\gev^3)$~&~$0.056\pm0.015$~&~$0.10\pm0.04$~&~$0.36$~&~$-0.238\pm0.121$~&~$0.059\pm0.019$ \\
$\langle{\cal O}^{\eta_c}(^3S_1^{[8]})\rangle(10^{-2}\gev^3)$~&~$3.04\pm0.35$~&~$8.9\pm0.98$&~$0.74$~&~$11.0\pm1.4$~&~$5.66\pm0.47$ \\
$\langle{\cal O}^{\eta_c}(^1P_1^{[8]})\rangle/m_c^2(10^{-2}\gev^3)$~&~$-1.21\pm0.21$~&~$1.68\pm0.63$~&~$6.0$~&~$-0.936\pm0.453$~&~$1.03\pm0.31$ \\
\hline
\hline
\end{tabular}
\end{center}
\end{table*}

\begin{figure}
\includegraphics[scale=0.5]{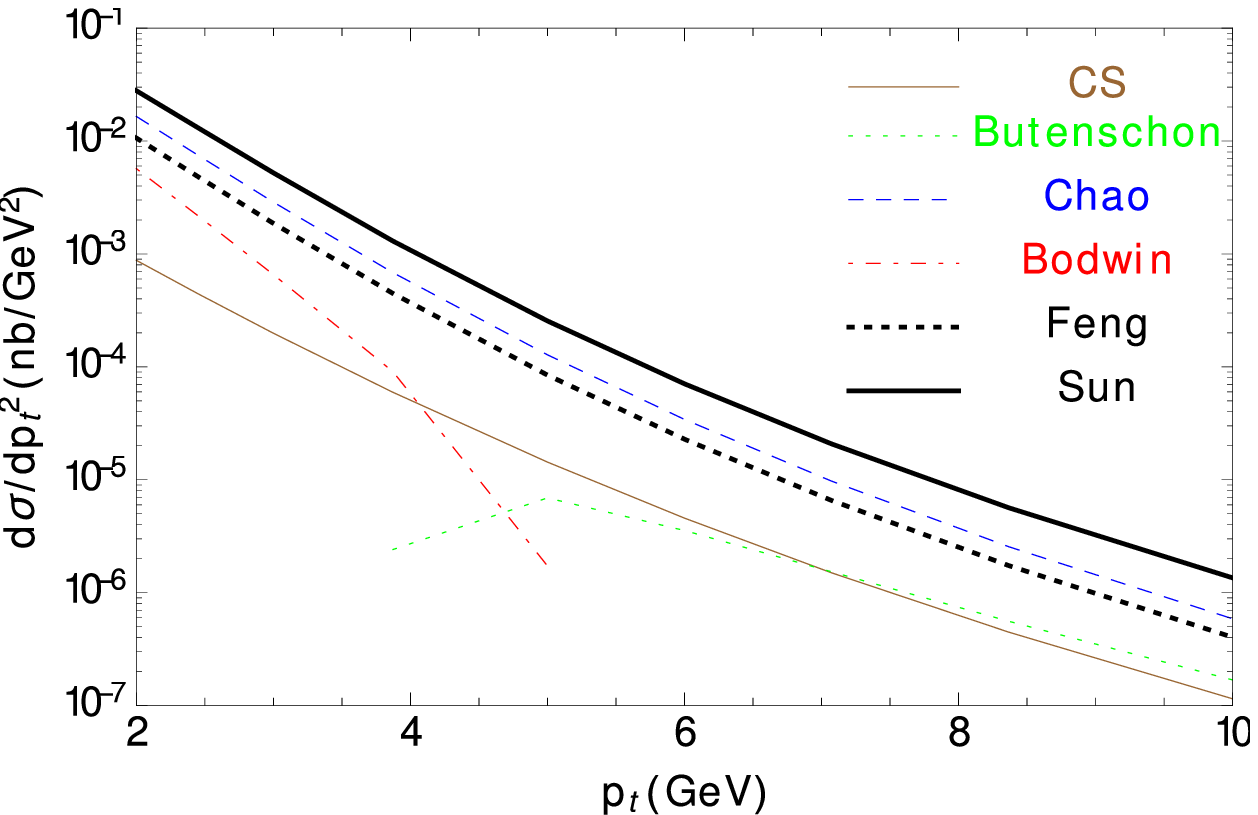}\\
\includegraphics[scale=0.5]{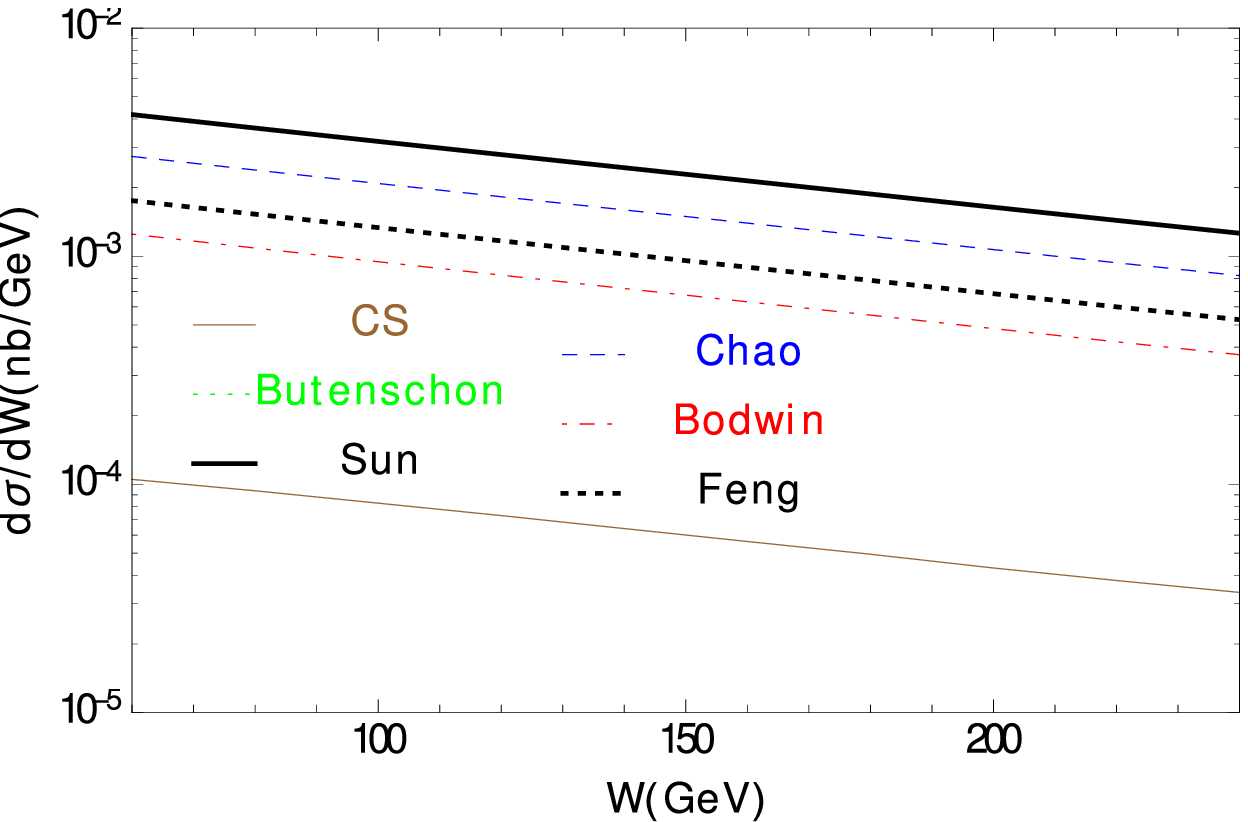}\\
\includegraphics[scale=0.5]{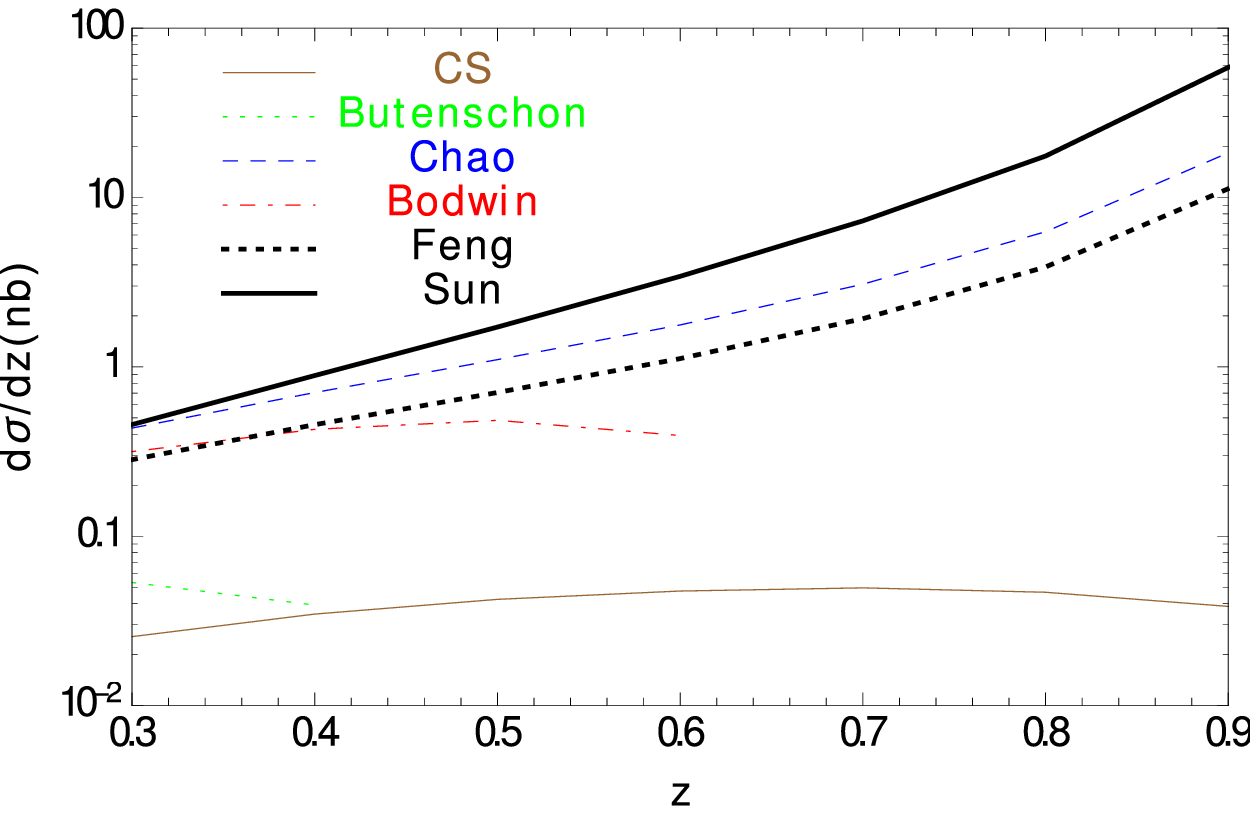}
\caption{\label{fig:ldme}
The $p_t$ (upper), $W$ (mid), and $z$ (lower) distribution of $\eta_c$ photoproduction employing different sets of LDMEs.
}
\end{figure}

The numerical results for different sets of LDMEs are given in Fig.~\ref{fig:ldme}.
Although these LDMEs lead to almost equal results for the $J/\psi$ hadroproduction,
their prediction on the $\eta_c$ photoproduction are completely different.
We can see that most of the curves are above the CS ones, except for those obtained by using the LDMEs given in Ref.~\cite{Butenschoen:2011yh},
which lead to negative differential cross sections in small $p_t$ region, large $z$ region, and the whole $W$ range.
With the LDMEs given in Ref.~\cite{Bodwin:2015iua}, we also get negative results in high $p_t$ and $z$ regions.
With the other sets of LDMEs, the CO results are positive and at least one order of magnitude larger than the CS onces.
These characteristic features in the $\eta_c$ production can provide an ideal laboratory to investigate the CO mechanism of the NRQCD,
as well as to scrutinize the validity of different sets of LDMEs.

\section{Summary\label{sec:summary}}

In this paper, we calculate the $\eta_c$ photoproduction in $ep$ collisions at QCD LO,
including not only the CO contributions, but also the CS ones for the first time.
Employing different sets of LDMEs, we find that the CO results are generally much larger than the CS ones.
This feature can be utilized to test the CO mechanism.
We also find that the different sets of LDMEs lead to different phenomenological results,
which enables this process to be an ideal laboratory to distinguish the variety of LDMEs on the market and impose new constraints on these LDMEs.
We suggest that the future electron-proton and electron-nucleus colliders could measure the $\eta_c$ production,
which may offer great help for the study of the quarkonium production mechanism.

\begin{acknowledgments}
This work is supported by the National Natural Science Foundation of China (Grant Nos. 11605144, 11747037).
Y.-P. Y. acknowledges support from SUT and the Office of the Higher Education Commission under the NRU project of Thailand.
\end{acknowledgments}

%\bibliography{paper}% Produces the bibliography via BibTeX.

\begin{thebibliography}{46}
\expandafter\ifx\csname natexlab\endcsname\relax\def\natexlab#1{#1}\fi
\expandafter\ifx\csname bibnamefont\endcsname\relax
  \def\bibnamefont#1{#1}\fi
\expandafter\ifx\csname bibfnamefont\endcsname\relax
  \def\bibfnamefont#1{#1}\fi
\expandafter\ifx\csname citenamefont\endcsname\relax
  \def\citenamefont#1{#1}\fi
\expandafter\ifx\csname url\endcsname\relax
  \def\url#1{\texttt{#1}}\fi
\expandafter\ifx\csname urlprefix\endcsname\relax\def\urlprefix{URL }\fi
\providecommand{\bibinfo}[2]{#2}
\providecommand{\eprint}[2][]{\url{#2}}

\bibitem[{\citenamefont{Barsuk et~al.}(2012)\citenamefont{Barsuk, He, Kou, and
  Viaud}}]{Barsuk:2012ic}
\bibinfo{author}{\bibfnamefont{S.}~\bibnamefont{Barsuk}},
  \bibinfo{author}{\bibfnamefont{J.}~\bibnamefont{He}},
  \bibinfo{author}{\bibfnamefont{E.}~\bibnamefont{Kou}}, \bibnamefont{and}
  \bibinfo{author}{\bibfnamefont{B.}~\bibnamefont{Viaud}},
  \bibinfo{journal}{Phys.Rev.} \textbf{\bibinfo{volume}{D86}},
  \bibinfo{pages}{034011} (\bibinfo{year}{2012}), \eprint{1202.2273}.

\bibitem[{\citenamefont{Aaij et~al.}(2015)}]{Aaij:2014bga}
\bibinfo{author}{\bibfnamefont{R.}~\bibnamefont{Aaij}} \bibnamefont{et~al.}
  (\bibinfo{collaboration}{LHCb}), \bibinfo{journal}{Eur. Phys. J.}
  \textbf{\bibinfo{volume}{C75}}, \bibinfo{pages}{311} (\bibinfo{year}{2015}),
  \eprint{1409.3612}.

\bibitem[{\citenamefont{Butenschoen et~al.}(2015)\citenamefont{Butenschoen, He,
  and Kniehl}}]{Butenschoen:2014dra}
\bibinfo{author}{\bibfnamefont{M.}~\bibnamefont{Butenschoen}},
  \bibinfo{author}{\bibfnamefont{Z.-G.} \bibnamefont{He}}, \bibnamefont{and}
  \bibinfo{author}{\bibfnamefont{B.~A.} \bibnamefont{Kniehl}},
  \bibinfo{journal}{Phys. Rev. Lett.} \textbf{\bibinfo{volume}{114}},
  \bibinfo{pages}{092004} (\bibinfo{year}{2015}), \eprint{1411.5287}.

\bibitem[{\citenamefont{Han et~al.}(2015)\citenamefont{Han, Ma, Meng, Shao, and
  Chao}}]{Han:2014jya}
\bibinfo{author}{\bibfnamefont{H.}~\bibnamefont{Han}},
  \bibinfo{author}{\bibfnamefont{Y.-Q.} \bibnamefont{Ma}},
  \bibinfo{author}{\bibfnamefont{C.}~\bibnamefont{Meng}},
  \bibinfo{author}{\bibfnamefont{H.-S.} \bibnamefont{Shao}}, \bibnamefont{and}
  \bibinfo{author}{\bibfnamefont{K.-T.} \bibnamefont{Chao}},
  \bibinfo{journal}{Phys. Rev. Lett.} \textbf{\bibinfo{volume}{114}},
  \bibinfo{pages}{092005} (\bibinfo{year}{2015}), \eprint{1411.7350}.

\bibitem[{\citenamefont{Zhang et~al.}(2015)\citenamefont{Zhang, Sun, Sang, and
  Li}}]{Zhang:2014ybe}
\bibinfo{author}{\bibfnamefont{H.-F.} \bibnamefont{Zhang}},
  \bibinfo{author}{\bibfnamefont{Z.}~\bibnamefont{Sun}},
  \bibinfo{author}{\bibfnamefont{W.-L.} \bibnamefont{Sang}}, \bibnamefont{and}
  \bibinfo{author}{\bibfnamefont{R.}~\bibnamefont{Li}}, \bibinfo{journal}{Phys.
  Rev. Lett.} \textbf{\bibinfo{volume}{114}}, \bibinfo{pages}{092006}
  (\bibinfo{year}{2015}), \eprint{1412.0508}.

\bibitem[{\citenamefont{Bodwin et~al.}(1995)\citenamefont{Bodwin, Braaten, and
  Lepage}}]{Bodwin:1994jh}
\bibinfo{author}{\bibfnamefont{G.~T.} \bibnamefont{Bodwin}},
  \bibinfo{author}{\bibfnamefont{E.}~\bibnamefont{Braaten}}, \bibnamefont{and}
  \bibinfo{author}{\bibfnamefont{G.~P.} \bibnamefont{Lepage}},
  \bibinfo{journal}{Phys. Rev.} \textbf{\bibinfo{volume}{D51}},
  \bibinfo{pages}{1125} (\bibinfo{year}{1995}), \bibinfo{note}{[Erratum: Phys.
  Rev.D55,5853(1997)]}, \eprint{hep-ph/9407339}.

\bibitem[{\citenamefont{Butenschoen et~al.}(2017)\citenamefont{Butenschoen, He,
  and Kniehl}}]{Butenschoen:2017iks}
\bibinfo{author}{\bibfnamefont{M.}~\bibnamefont{Butenschoen}},
  \bibinfo{author}{\bibfnamefont{Z.-G.} \bibnamefont{He}}, \bibnamefont{and}
  \bibinfo{author}{\bibfnamefont{B.~A.} \bibnamefont{Kniehl}},
  \bibinfo{journal}{EPJ Web Conf.} \textbf{\bibinfo{volume}{137}},
  \bibinfo{pages}{06009} (\bibinfo{year}{2017}).

\bibitem[{\citenamefont{Feng et~al.}(2015)\citenamefont{Feng, Jia, and
  Sang}}]{Feng:2015uha}
\bibinfo{author}{\bibfnamefont{F.}~\bibnamefont{Feng}},
  \bibinfo{author}{\bibfnamefont{Y.}~\bibnamefont{Jia}}, \bibnamefont{and}
  \bibinfo{author}{\bibfnamefont{W.-L.} \bibnamefont{Sang}},
  \bibinfo{journal}{Phys. Rev. Lett.} \textbf{\bibinfo{volume}{115}},
  \bibinfo{pages}{222001} (\bibinfo{year}{2015}), \eprint{1505.02665}.

\bibitem[{\citenamefont{Sang et~al.}(2016)\citenamefont{Sang, Feng, Jia, and
  Liang}}]{Sang:2015uxg}
\bibinfo{author}{\bibfnamefont{W.-L.} \bibnamefont{Sang}},
  \bibinfo{author}{\bibfnamefont{F.}~\bibnamefont{Feng}},
  \bibinfo{author}{\bibfnamefont{Y.}~\bibnamefont{Jia}}, \bibnamefont{and}
  \bibinfo{author}{\bibfnamefont{S.-R.} \bibnamefont{Liang}},
  \bibinfo{journal}{Phys. Rev.} \textbf{\bibinfo{volume}{D94}},
  \bibinfo{pages}{111501} (\bibinfo{year}{2016}), \eprint{1511.06288}.

\bibitem[{\citenamefont{Chen et~al.}(2017)\citenamefont{Chen, Liang, and
  Qiao}}]{Chen:2017xqd}
\bibinfo{author}{\bibfnamefont{L.-B.} \bibnamefont{Chen}},
  \bibinfo{author}{\bibfnamefont{Y.}~\bibnamefont{Liang}}, \bibnamefont{and}
  \bibinfo{author}{\bibfnamefont{C.-F.} \bibnamefont{Qiao}},
  \bibinfo{journal}{JHEP} \textbf{\bibinfo{volume}{06}}, \bibinfo{pages}{025}
  (\bibinfo{year}{2017}), \eprint{1703.03929}.

\bibitem[{\citenamefont{Feng et~al.}(2017)\citenamefont{Feng, Jia, and
  Sang}}]{Feng:2017hlu}
\bibinfo{author}{\bibfnamefont{F.}~\bibnamefont{Feng}},
  \bibinfo{author}{\bibfnamefont{Y.}~\bibnamefont{Jia}}, \bibnamefont{and}
  \bibinfo{author}{\bibfnamefont{W.-L.} \bibnamefont{Sang}},
  \bibinfo{journal}{Phys. Rev. Lett.} \textbf{\bibinfo{volume}{119}},
  \bibinfo{pages}{252001} (\bibinfo{year}{2017}), \eprint{1707.05758}.

\bibitem[{\citenamefont{Chen et~al.}(2018{\natexlab{a}})\citenamefont{Chen,
  Liang, and Qiao}}]{Chen:2017pyi}
\bibinfo{author}{\bibfnamefont{L.-B.} \bibnamefont{Chen}},
  \bibinfo{author}{\bibfnamefont{Y.}~\bibnamefont{Liang}}, \bibnamefont{and}
  \bibinfo{author}{\bibfnamefont{C.-F.} \bibnamefont{Qiao}},
  \bibinfo{journal}{JHEP} \textbf{\bibinfo{volume}{01}}, \bibinfo{pages}{091}
  (\bibinfo{year}{2018}{\natexlab{a}}), \eprint{1710.07865}.

\bibitem[{\citenamefont{Chen et~al.}(2018{\natexlab{b}})\citenamefont{Chen,
  Jiang, and Qiao}}]{Chen:2017soz}
\bibinfo{author}{\bibfnamefont{L.-B.} \bibnamefont{Chen}},
  \bibinfo{author}{\bibfnamefont{J.}~\bibnamefont{Jiang}}, \bibnamefont{and}
  \bibinfo{author}{\bibfnamefont{C.-F.} \bibnamefont{Qiao}},
  \bibinfo{journal}{JHEP} \textbf{\bibinfo{volume}{04}}, \bibinfo{pages}{080}
  (\bibinfo{year}{2018}{\natexlab{b}}), \eprint{1712.03516}.

\bibitem[{\citenamefont{Feng et~al.}(2019{\natexlab{a}})\citenamefont{Feng,
  Jia, and Sang}}]{Feng:2019zmt}
\bibinfo{author}{\bibfnamefont{F.}~\bibnamefont{Feng}},
  \bibinfo{author}{\bibfnamefont{Y.}~\bibnamefont{Jia}}, \bibnamefont{and}
  \bibinfo{author}{\bibfnamefont{W.-L.} \bibnamefont{Sang}}
  (\bibinfo{year}{2019}{\natexlab{a}}), \eprint{1901.08447}.

\bibitem[{\citenamefont{Ma and Chao}(2017)}]{Ma:2017xno}
\bibinfo{author}{\bibfnamefont{Y.-Q.} \bibnamefont{Ma}} \bibnamefont{and}
  \bibinfo{author}{\bibfnamefont{K.-T.} \bibnamefont{Chao}}
  (\bibinfo{year}{2017}), \eprint{1703.08402}.

\bibitem[{\citenamefont{Bodwin et~al.}(2014)\citenamefont{Bodwin, Chung, Kim,
  and Lee}}]{Bodwin:2014gia}
\bibinfo{author}{\bibfnamefont{G.~T.} \bibnamefont{Bodwin}},
  \bibinfo{author}{\bibfnamefont{H.~S.} \bibnamefont{Chung}},
  \bibinfo{author}{\bibfnamefont{U.-R.} \bibnamefont{Kim}}, \bibnamefont{and}
  \bibinfo{author}{\bibfnamefont{J.}~\bibnamefont{Lee}},
  \bibinfo{journal}{Phys. Rev. Lett.} \textbf{\bibinfo{volume}{113}},
  \bibinfo{pages}{022001} (\bibinfo{year}{2014}), \eprint{1403.3612}.

\bibitem[{\citenamefont{Bodwin et~al.}(2015)\citenamefont{Bodwin, Chung, Kim,
  and Lee}}]{Bodwin:2015yma}
\bibinfo{author}{\bibfnamefont{G.~T.} \bibnamefont{Bodwin}},
  \bibinfo{author}{\bibfnamefont{H.~S.} \bibnamefont{Chung}},
  \bibinfo{author}{\bibfnamefont{U.-R.} \bibnamefont{Kim}}, \bibnamefont{and}
  \bibinfo{author}{\bibfnamefont{J.}~\bibnamefont{Lee}},
  \bibinfo{journal}{Phys. Rev.} \textbf{\bibinfo{volume}{D92}},
  \bibinfo{pages}{074042} (\bibinfo{year}{2015}), \eprint{1504.06019}.

\bibitem[{\citenamefont{Bodwin et~al.}(2016)\citenamefont{Bodwin, Chao, Chung,
  Kim, Lee, and Ma}}]{Bodwin:2015iua}
\bibinfo{author}{\bibfnamefont{G.~T.} \bibnamefont{Bodwin}},
  \bibinfo{author}{\bibfnamefont{K.-T.} \bibnamefont{Chao}},
  \bibinfo{author}{\bibfnamefont{H.~S.} \bibnamefont{Chung}},
  \bibinfo{author}{\bibfnamefont{U.-R.} \bibnamefont{Kim}},
  \bibinfo{author}{\bibfnamefont{J.}~\bibnamefont{Lee}}, \bibnamefont{and}
  \bibinfo{author}{\bibfnamefont{Y.-Q.} \bibnamefont{Ma}},
  \bibinfo{journal}{Phys. Rev.} \textbf{\bibinfo{volume}{D93}},
  \bibinfo{pages}{034041} (\bibinfo{year}{2016}), \eprint{1509.07904}.

\bibitem[{\citenamefont{Campbell et~al.}(2007)\citenamefont{Campbell, Maltoni,
  and Tramontano}}]{Campbell:2007ws}
\bibinfo{author}{\bibfnamefont{J.~M.} \bibnamefont{Campbell}},
  \bibinfo{author}{\bibfnamefont{F.}~\bibnamefont{Maltoni}}, \bibnamefont{and}
  \bibinfo{author}{\bibfnamefont{F.}~\bibnamefont{Tramontano}},
  \bibinfo{journal}{Phys. Rev. Lett.} \textbf{\bibinfo{volume}{98}},
  \bibinfo{pages}{252002} (\bibinfo{year}{2007}), \eprint{hep-ph/0703113}.

\bibitem[{\citenamefont{Ma et~al.}(2011{\natexlab{a}})\citenamefont{Ma, Wang,
  and Chao}}]{Ma:2010yw}
\bibinfo{author}{\bibfnamefont{Y.-Q.} \bibnamefont{Ma}},
  \bibinfo{author}{\bibfnamefont{K.}~\bibnamefont{Wang}}, \bibnamefont{and}
  \bibinfo{author}{\bibfnamefont{K.-T.} \bibnamefont{Chao}},
  \bibinfo{journal}{Phys. Rev. Lett.} \textbf{\bibinfo{volume}{106}},
  \bibinfo{pages}{042002} (\bibinfo{year}{2011}{\natexlab{a}}),
  \eprint{1009.3655}.

\bibitem[{\citenamefont{Butenschoen and
  Kniehl}(2011{\natexlab{a}})}]{Butenschoen:2010rq}
\bibinfo{author}{\bibfnamefont{M.}~\bibnamefont{Butenschoen}} \bibnamefont{and}
  \bibinfo{author}{\bibfnamefont{B.~A.} \bibnamefont{Kniehl}},
  \bibinfo{journal}{Phys. Rev. Lett.} \textbf{\bibinfo{volume}{106}},
  \bibinfo{pages}{022003} (\bibinfo{year}{2011}{\natexlab{a}}),
  \eprint{1009.5662}.

\bibitem[{\citenamefont{Ma et~al.}(2011{\natexlab{b}})\citenamefont{Ma, Wang,
  and Chao}}]{Ma:2010jj}
\bibinfo{author}{\bibfnamefont{Y.-Q.} \bibnamefont{Ma}},
  \bibinfo{author}{\bibfnamefont{K.}~\bibnamefont{Wang}}, \bibnamefont{and}
  \bibinfo{author}{\bibfnamefont{K.-T.} \bibnamefont{Chao}},
  \bibinfo{journal}{Phys. Rev.} \textbf{\bibinfo{volume}{D84}},
  \bibinfo{pages}{114001} (\bibinfo{year}{2011}{\natexlab{b}}),
  \eprint{1012.1030}.

\bibitem[{\citenamefont{Butenschoen and
  Kniehl}(2011{\natexlab{b}})}]{Butenschoen:2011yh}
\bibinfo{author}{\bibfnamefont{M.}~\bibnamefont{Butenschoen}} \bibnamefont{and}
  \bibinfo{author}{\bibfnamefont{B.~A.} \bibnamefont{Kniehl}},
  \bibinfo{journal}{Phys. Rev.} \textbf{\bibinfo{volume}{D84}},
  \bibinfo{pages}{051501} (\bibinfo{year}{2011}{\natexlab{b}}),
  \eprint{1105.0820}.

\bibitem[{\citenamefont{Butenschoen and Kniehl}(2012)}]{Butenschoen:2012px}
\bibinfo{author}{\bibfnamefont{M.}~\bibnamefont{Butenschoen}} \bibnamefont{and}
  \bibinfo{author}{\bibfnamefont{B.~A.} \bibnamefont{Kniehl}},
  \bibinfo{journal}{Phys. Rev. Lett.} \textbf{\bibinfo{volume}{108}},
  \bibinfo{pages}{172002} (\bibinfo{year}{2012}), \eprint{1201.1872}.

\bibitem[{\citenamefont{Chao et~al.}(2012)\citenamefont{Chao, Ma, Shao, Wang,
  and Zhang}}]{Chao:2012iv}
\bibinfo{author}{\bibfnamefont{K.-T.} \bibnamefont{Chao}},
  \bibinfo{author}{\bibfnamefont{Y.-Q.} \bibnamefont{Ma}},
  \bibinfo{author}{\bibfnamefont{H.-S.} \bibnamefont{Shao}},
  \bibinfo{author}{\bibfnamefont{K.}~\bibnamefont{Wang}}, \bibnamefont{and}
  \bibinfo{author}{\bibfnamefont{Y.-J.} \bibnamefont{Zhang}},
  \bibinfo{journal}{Phys. Rev. Lett.} \textbf{\bibinfo{volume}{108}},
  \bibinfo{pages}{242004} (\bibinfo{year}{2012}), \eprint{1201.2675}.

\bibitem[{\citenamefont{Gong et~al.}(2013)\citenamefont{Gong, Wan, Wang, and
  Zhang}}]{Gong:2012ug}
\bibinfo{author}{\bibfnamefont{B.}~\bibnamefont{Gong}},
  \bibinfo{author}{\bibfnamefont{L.-P.} \bibnamefont{Wan}},
  \bibinfo{author}{\bibfnamefont{J.-X.} \bibnamefont{Wang}}, \bibnamefont{and}
  \bibinfo{author}{\bibfnamefont{H.-F.} \bibnamefont{Zhang}},
  \bibinfo{journal}{Phys. Rev. Lett.} \textbf{\bibinfo{volume}{110}},
  \bibinfo{pages}{042002} (\bibinfo{year}{2013}), \eprint{1205.6682}.

\bibitem[{\citenamefont{Shao et~al.}(2015)\citenamefont{Shao, Han, Ma, Meng,
  Zhang, and Chao}}]{Shao:2014yta}
\bibinfo{author}{\bibfnamefont{H.~S.} \bibnamefont{Shao}},
  \bibinfo{author}{\bibfnamefont{H.}~\bibnamefont{Han}},
  \bibinfo{author}{\bibfnamefont{Y.~Q.} \bibnamefont{Ma}},
  \bibinfo{author}{\bibfnamefont{C.}~\bibnamefont{Meng}},
  \bibinfo{author}{\bibfnamefont{Y.~J.} \bibnamefont{Zhang}}, \bibnamefont{and}
  \bibinfo{author}{\bibfnamefont{K.~T.} \bibnamefont{Chao}},
  \bibinfo{journal}{JHEP} \textbf{\bibinfo{volume}{05}}, \bibinfo{pages}{103}
  (\bibinfo{year}{2015}), \eprint{1411.3300}.

\bibitem[{\citenamefont{Sun and Zhang}(2018)}]{Sun:2015pia}
\bibinfo{author}{\bibfnamefont{Z.}~\bibnamefont{Sun}} \bibnamefont{and}
  \bibinfo{author}{\bibfnamefont{H.-F.} \bibnamefont{Zhang}},
  \bibinfo{journal}{Chin. Phys.} \textbf{\bibinfo{volume}{C42}},
  \bibinfo{pages}{043104} (\bibinfo{year}{2018}), \eprint{1505.02675}.

\bibitem[{\citenamefont{Feng et~al.}(2019{\natexlab{b}})\citenamefont{Feng,
  Gong, Chang, and Wang}}]{Feng:2018ukp}
\bibinfo{author}{\bibfnamefont{Y.}~\bibnamefont{Feng}},
  \bibinfo{author}{\bibfnamefont{B.}~\bibnamefont{Gong}},
  \bibinfo{author}{\bibfnamefont{C.-H.} \bibnamefont{Chang}}, \bibnamefont{and}
  \bibinfo{author}{\bibfnamefont{J.-X.} \bibnamefont{Wang}},
  \bibinfo{journal}{Phys. Rev.} \textbf{\bibinfo{volume}{D99}},
  \bibinfo{pages}{014044} (\bibinfo{year}{2019}{\natexlab{b}}),
  \eprint{1810.08989}.

\bibitem[{\citenamefont{Artoisenet et~al.}(2009)\citenamefont{Artoisenet,
  Campbell, Maltoni, and Tramontano}}]{Artoisenet:2009xh}
\bibinfo{author}{\bibfnamefont{P.}~\bibnamefont{Artoisenet}},
  \bibinfo{author}{\bibfnamefont{J.~M.} \bibnamefont{Campbell}},
  \bibinfo{author}{\bibfnamefont{F.}~\bibnamefont{Maltoni}}, \bibnamefont{and}
  \bibinfo{author}{\bibfnamefont{F.}~\bibnamefont{Tramontano}},
  \bibinfo{journal}{Phys. Rev. Lett.} \textbf{\bibinfo{volume}{102}},
  \bibinfo{pages}{142001} (\bibinfo{year}{2009}), \eprint{0901.4352}.

\bibitem[{\citenamefont{Butenschoen and Kniehl}(2010)}]{Butenschoen:2009zy}
\bibinfo{author}{\bibfnamefont{M.}~\bibnamefont{Butenschoen}} \bibnamefont{and}
  \bibinfo{author}{\bibfnamefont{B.~A.} \bibnamefont{Kniehl}},
  \bibinfo{journal}{Phys. Rev. Lett.} \textbf{\bibinfo{volume}{104}},
  \bibinfo{pages}{072001} (\bibinfo{year}{2010}), \eprint{0909.2798}.

\bibitem[{\citenamefont{Butenschoen and
  Kniehl}(2011{\natexlab{c}})}]{Butenschoen:2011ks}
\bibinfo{author}{\bibfnamefont{M.}~\bibnamefont{Butenschoen}} \bibnamefont{and}
  \bibinfo{author}{\bibfnamefont{B.~A.} \bibnamefont{Kniehl}},
  \bibinfo{journal}{Phys. Rev. Lett.} \textbf{\bibinfo{volume}{107}},
  \bibinfo{pages}{232001} (\bibinfo{year}{2011}{\natexlab{c}}),
  \eprint{1109.1476}.

\bibitem[{\citenamefont{Sun and Zhang}(2017{\natexlab{a}})}]{Sun:2017wxk}
\bibinfo{author}{\bibfnamefont{Z.}~\bibnamefont{Sun}} \bibnamefont{and}
  \bibinfo{author}{\bibfnamefont{H.-F.} \bibnamefont{Zhang}},
  \bibinfo{journal}{Phys. Rev.} \textbf{\bibinfo{volume}{D96}},
  \bibinfo{pages}{091502} (\bibinfo{year}{2017}{\natexlab{a}}),
  \eprint{1705.05337}.

\bibitem[{\citenamefont{Zhang and Sun}(2017)}]{Zhang:2017dia}
\bibinfo{author}{\bibfnamefont{H.-F.} \bibnamefont{Zhang}} \bibnamefont{and}
  \bibinfo{author}{\bibfnamefont{Z.}~\bibnamefont{Sun}},
  \bibinfo{journal}{Phys. Rev.} \textbf{\bibinfo{volume}{D96}},
  \bibinfo{pages}{034002} (\bibinfo{year}{2017}), \eprint{1701.08728}.

\bibitem[{\citenamefont{Sun and Zhang}(2017{\natexlab{b}})}]{Sun:2017nly}
\bibinfo{author}{\bibfnamefont{Z.}~\bibnamefont{Sun}} \bibnamefont{and}
  \bibinfo{author}{\bibfnamefont{H.-F.} \bibnamefont{Zhang}},
  \bibinfo{journal}{Eur. Phys. J.} \textbf{\bibinfo{volume}{C77}},
  \bibinfo{pages}{744} (\bibinfo{year}{2017}{\natexlab{b}}),
  \eprint{1702.02097}.

\bibitem[{\citenamefont{Gong et~al.}(2016)\citenamefont{Gong, Sun, Zhang, and
  Mo}}]{Gong:2016jiq}
\bibinfo{author}{\bibfnamefont{Q.-R.} \bibnamefont{Gong}},
  \bibinfo{author}{\bibfnamefont{Z.}~\bibnamefont{Sun}},
  \bibinfo{author}{\bibfnamefont{H.-F.} \bibnamefont{Zhang}}, \bibnamefont{and}
  \bibinfo{author}{\bibfnamefont{X.-M.} \bibnamefont{Mo}},
  \bibinfo{journal}{Eur. Phys. J.} \textbf{\bibinfo{volume}{C76}},
  \bibinfo{pages}{518} (\bibinfo{year}{2016}), \eprint{1606.08317}.

\bibitem[{\citenamefont{Lansberg et~al.}(2018)\citenamefont{Lansberg, Shao, and
  Zhang}}]{Lansberg:2017ozx}
\bibinfo{author}{\bibfnamefont{J.-P.} \bibnamefont{Lansberg}},
  \bibinfo{author}{\bibfnamefont{H.-S.} \bibnamefont{Shao}}, \bibnamefont{and}
  \bibinfo{author}{\bibfnamefont{H.-F.} \bibnamefont{Zhang}},
  \bibinfo{journal}{Phys. Lett.} \textbf{\bibinfo{volume}{B786}},
  \bibinfo{pages}{342} (\bibinfo{year}{2018}), \eprint{1711.00265}.

\bibitem[{\citenamefont{Feng et~al.}(2019{\natexlab{c}})\citenamefont{Feng, He,
  Lansberg, Shao, Usachov, and Zhang}}]{Feng:2019zmn}
\bibinfo{author}{\bibfnamefont{Y.}~\bibnamefont{Feng}},
  \bibinfo{author}{\bibfnamefont{J.}~\bibnamefont{He}},
  \bibinfo{author}{\bibfnamefont{J.-P.} \bibnamefont{Lansberg}},
  \bibinfo{author}{\bibfnamefont{H.-S.} \bibnamefont{Shao}},
  \bibinfo{author}{\bibfnamefont{A.}~\bibnamefont{Usachov}}, \bibnamefont{and}
  \bibinfo{author}{\bibfnamefont{H.-F.} \bibnamefont{Zhang}}
  (\bibinfo{year}{2019}{\natexlab{c}}), \eprint{1901.09766}.

\bibitem[{\citenamefont{Hao et~al.}(1999)\citenamefont{Hao, Yuan, and
  Chao}}]{Hao:1999kq}
\bibinfo{author}{\bibfnamefont{L.-K.} \bibnamefont{Hao}},
  \bibinfo{author}{\bibfnamefont{F.}~\bibnamefont{Yuan}}, \bibnamefont{and}
  \bibinfo{author}{\bibfnamefont{K.-T.} \bibnamefont{Chao}},
  \bibinfo{journal}{Phys. Rev. Lett.} \textbf{\bibinfo{volume}{83}},
  \bibinfo{pages}{4490} (\bibinfo{year}{1999}), \eprint{hep-ph/9902338}.

\bibitem[{\citenamefont{Hao et~al.}(2000)\citenamefont{Hao, Yuan, and
  Chao}}]{Hao:2000ci}
\bibinfo{author}{\bibfnamefont{L.-K.} \bibnamefont{Hao}},
  \bibinfo{author}{\bibfnamefont{F.}~\bibnamefont{Yuan}}, \bibnamefont{and}
  \bibinfo{author}{\bibfnamefont{K.-T.} \bibnamefont{Chao}},
  \bibinfo{journal}{Phys. Rev.} \textbf{\bibinfo{volume}{D62}},
  \bibinfo{pages}{074023} (\bibinfo{year}{2000}), \eprint{hep-ph/0004203}.

\bibitem[{\citenamefont{von Weizsacker}(1934)}]{vonWeizsacker:1934nji}
\bibinfo{author}{\bibfnamefont{C.~F.} \bibnamefont{von Weizsacker}},
  \bibinfo{journal}{Z. Phys.} \textbf{\bibinfo{volume}{88}},
  \bibinfo{pages}{612} (\bibinfo{year}{1934}).

\bibitem[{\citenamefont{Williams}(1934)}]{Williams:1934ad}
\bibinfo{author}{\bibfnamefont{E.~J.} \bibnamefont{Williams}},
  \bibinfo{journal}{Phys. Rev.} \textbf{\bibinfo{volume}{45}},
  \bibinfo{pages}{729} (\bibinfo{year}{1934}).

\bibitem[{\citenamefont{Frixione et~al.}(1993)\citenamefont{Frixione, Mangano,
  Nason, and Ridolfi}}]{Frixione:1993yw}
\bibinfo{author}{\bibfnamefont{S.}~\bibnamefont{Frixione}},
  \bibinfo{author}{\bibfnamefont{M.~L.} \bibnamefont{Mangano}},
  \bibinfo{author}{\bibfnamefont{P.}~\bibnamefont{Nason}}, \bibnamefont{and}
  \bibinfo{author}{\bibfnamefont{G.}~\bibnamefont{Ridolfi}},
  \bibinfo{journal}{Phys. Lett.} \textbf{\bibinfo{volume}{B319}},
  \bibinfo{pages}{339} (\bibinfo{year}{1993}), \eprint{hep-ph/9310350}.

\bibitem[{\citenamefont{Wang}(2004)}]{Wang:2004du}
\bibinfo{author}{\bibfnamefont{J.-X.} \bibnamefont{Wang}},
  \bibinfo{journal}{Nucl. Instrum. Meth.} \textbf{\bibinfo{volume}{A534}},
  \bibinfo{pages}{241} (\bibinfo{year}{2004}), \eprint{hep-ph/0407058}.

\bibitem[{\citenamefont{Pumplin et~al.}(2002)\citenamefont{Pumplin, Stump,
  Huston, Lai, Nadolsky, and Tung}}]{Pumplin:2002vw}
\bibinfo{author}{\bibfnamefont{J.}~\bibnamefont{Pumplin}},
  \bibinfo{author}{\bibfnamefont{D.~R.} \bibnamefont{Stump}},
  \bibinfo{author}{\bibfnamefont{J.}~\bibnamefont{Huston}},
  \bibinfo{author}{\bibfnamefont{H.~L.} \bibnamefont{Lai}},
  \bibinfo{author}{\bibfnamefont{P.~M.} \bibnamefont{Nadolsky}},
  \bibnamefont{and} \bibinfo{author}{\bibfnamefont{W.~K.} \bibnamefont{Tung}},
  \bibinfo{journal}{JHEP} \textbf{\bibinfo{volume}{07}}, \bibinfo{pages}{012}
  (\bibinfo{year}{2002}), \eprint{hep-ph/0201195}.

\bibitem[{\citenamefont{Eichten and Quigg}(1995)}]{Eichten:1995ch}
\bibinfo{author}{\bibfnamefont{E.~J.} \bibnamefont{Eichten}} \bibnamefont{and}
  \bibinfo{author}{\bibfnamefont{C.}~\bibnamefont{Quigg}},
  \bibinfo{journal}{Phys.Rev.} \textbf{\bibinfo{volume}{D52}},
  \bibinfo{pages}{1726} (\bibinfo{year}{1995}), \eprint{hep-ph/9503356}.

\end{thebibliography}

\end{document}